# Three-dimensional Printing of Complex Graphite Structures

Seyed Mohammad Sajadi[1]*, Shayan Enayat[2], Lívia Vásárhelyi[3], Alessandro Alabastri[4], Minghe Lou[5], Lucas M. Sassi[1], Alex Kutana[1], Sanjit Bhowmick[6], Christian Durante[1], Ákos Kukovecz[3], Anand B. Puthirath[1], Zoltán Kónya[3], Robert Vajtai[1], Peter Boul[7], Chandra Shekhar Tiwary[8]*, Muhammad M. Rahman[1]*, Pulickel M. Ajayan[1]*

[1]Department of Materials Science and NanoEngineering, Rice University, Texas, USA
[2]Department of Chemical and Biomolecular Engineering, Rice University, Texas, USA
[3]University of Szeged, Interdisciplinary Excellence Centre, Department of Applied and Environmental Chemistry, Szeged, Hungary
[4]Department of Electrical and Computer Engineering, Rice University, Texas, USA
[5]Department of Chemistry, Rice University, Texas, USA
[6]Bruker Nano Surfaces, Minneapolis, Minnesota, USA
[7]Aramco Americas, Texas, USA
[8]Metallurgical and Materials Engineering, Indian Institute of Technology Kharagpur, India

*Corresponds to: msajadi@rice.edu; Chandra.tiwary@metal.iitkgp.ac.in; mr64@rice.edu; ajayan@rice.edu

**Graphite, with many industrial applications, is one of the widely sought-after allotropes of carbon. The sp$^2$ hybridized and thermodynamically stable form of carbon forms a layered structure with strong in-plane carbon bonds and weak inter-layer van der Waals bonding. Graphite is also a high-temperature ceramic, and shaping them into complex geometries is challenging, given its limited sintering behavior even at high temperatures. Although the geometric design of the graphite structure in many of the applications could dictate its precision performance, conventional synthesis methods for formulating complex geometric graphite shapes are limited due to the intrinsic brittleness and difficulties of high-temperature processing. Here, we report the development of colloidal graphite ink from commercial graphite powders with reproducible rheological behavior that allows the fabrication of any complex architectures with tunable geometry and directionality via 3D printing at room temperature. The method is enabled via using small amounts of clay, another layered material, as an additive, allowing the proper design of the graphene ink and subsequent binding of graphite platelets during printing. Sheared layers of clay are easily able to flow, adapt, and interface with graphite layers forming strong binding between the layers and between particles that make the larger structures. The direct ink printing of complex 3D architectures of graphite without further heat treatments could lead to easy shape engineering and related applications of graphite at various length scales, including complex graphite molds or crucibles. The 3D printed complex graphitic structures exhibit excellent thermal, electrical, and mechanical properties, and the clay additive does not seem to alter these properties due to the excellent inter-layer dispersion and mixing within the graphite material.**

Graphite is one of the widely sought after and highly mined natural allotropes of carbon. It consists of superimposed lamellae of two-dimensional (2D) carbon-carbon covalent networks stacked along the *c*-axis via strong van der Waals forces. It is an ideal building block of three dimensional (3D) structures for a myriad of applications in many aggressive environments, including electrodes and electrical contact, catalyst support, gas adsorption, casting mold, nuclear reactors, heating elements, lab crucibles, and seals owing to its high electrical and thermal conductivity, minimal thermal expansion, and excellent chemical and thermal stability.[1–9] Among the worldwide graphite usage, the majority fraction (~40%) is consumed in refractory foundries as mold or high-temperature component, followed by (~30%) utilization in metallurgical industries as electrical contact/electrode material.[10] Furthermore, its specific deployment in the field of electronics and energy sectors continues to grow exponentially. In many of these applications, the geometric design of the graphitic structure determines its precision performance. Nevertheless, conventional synthesis methods have been limited in formulating the desired architecture due to the inherent brittleness and high processing temperature of natural graphite.[11] In theory, atomically thin graphene provides an opportunity to build a graphite block using the bottom-up approach, which would lower the processing temperature and enhance the strength of the material.[12,13] However, no practical method has yet been found to achieve such precise control over the directionality and plasticity of the restructuring process.

Patterning graphite through three-dimensional (3D) printing has the potential to enable the realization of architected structures with complex topology. The precision of 3D printing of graphite can open up a new perspective to create architected materials and composites with greater control over its directionality and local compositional requirement i.e., plasticity, which remains impossible to fabricate by the conventional manufacturing methods. Among the 3D printing methods, direct ink writing (DIW), an extrusion-based process, offers rapid fabrication of complex shapes by deposition of colloidal inks in a layer-by-layer fashion. This technique allows excellent compatibility with a wide array of materials, including polymers, ceramics, cement, and composites, to develop 3D structures with outstanding properties and multifunctionality.[14,15] However, the design of a colloidal graphite ink with critical reproducible rheology still remains a challenge. The graphite ink must exhibit shear-thinning behavior with desired apparent viscosity, which facilitates the extrusion of the ink through the nozzle without high printing pressure. Besides, the printing ink must be integrated with appropriate viscoelastic properties such as high storage modulus and yield strength to maintain its filamentary shape after extrusion from the nozzle. As such, the controlled assembly of graphite into tailored 3D architectures via DIW has not been reported.

Herein for the first time, we report the development of colloidal graphite ink that allows fabrication of 3D complex architectures with tunable geometry and directionality in ambient conditions. The graphite ink has successfully addressed the essential rheological characteristics such as shear thinning and rapid gel strength for DIW of architected structure. The shear-induced alignment of the graphite sheet during extrusion of the ink was observed along the printing direction through a Scanning Electron Microscope (SEM), and a simulation of the flow of ink through the nozzle was implemented. Structural characterization of the printed graphite architecture was conducted using X-ray diffraction (XRD), Raman spectroscopy, and X-ray photoelectron spectroscopy (XPS). High-resolution spatial and temporal mapping of temperature variations during irradiation, four-point probe current-voltage (IV) measurement, and uniaxial compression tests were performed for thermal, electrical, and mechanical characterization of the

3D printed structure, respectively. We also demonstrate the potential application of the printed structures in complex metal mold casting, heating element, and electrical circuit/electrode.

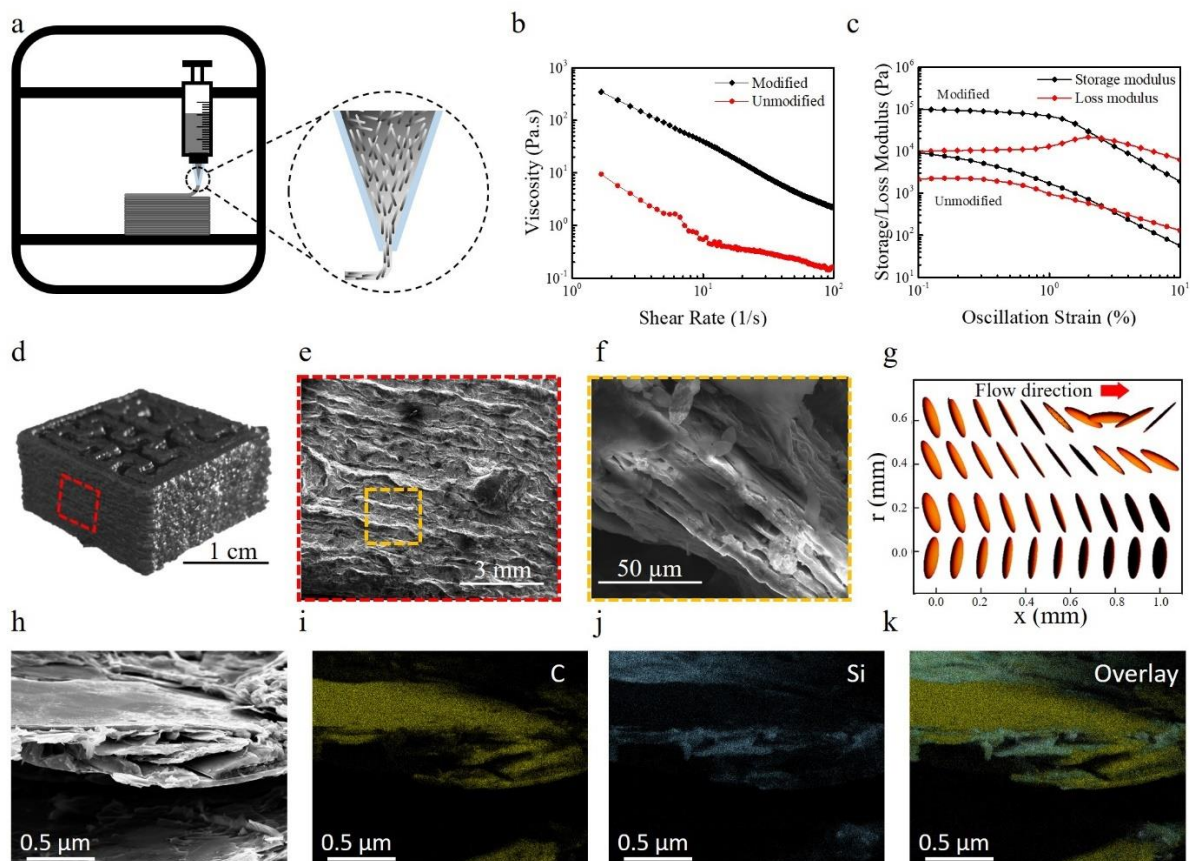

**Figure 1. 3D printing of graphite.** a) Schematic representation of 3D printing of graphite through direct ink writing. Higher magnification (schematic) demonstrates how the graphite sheets align through the nozzle as a result of shear force. b) Viscosity vs. shear rate plot of the modified and unmodified graphite ink. The viscosity of modified ink is two orders of magnitude higher than unmodified ink and shows shear-thinning behavior. c) The storage and loss modulus of graphite inks as a function of oscillatory strain. The modified graphite ink has considerably higher storage modulus compared to the unmodified one. As a result of higher storage modulus, the printed structures maintain their structural integrity immediately after printing. d) Photograph of complex 3D printed graphite structure from an isometric perspective. e) SEM images from the side of the printed structure which illustrates the resolution of printing in the z-direction. Each layer has an approximate thickness of 500 μm. f) The alignment of graphite sheets in higher magnification. g) Alignment of graphite sheets with an aspect ratio (l) of 0.2 in a Poiseuille flow radius (r) of 0.8 mm and an average velocity of 12 mm/s (Figure shows half of the nozzle). The maximum alignment is near the wall where shear is highest. This plot represents half of the nozzle because of symmetrical boundary conditions. h) SEM micrograph of graphite sheet under higher magnification. i-k) Energy dispersive X-ray analysis, which demonstrates the location of carbon, silicon, and overlay (respectively) of these elements.

The key factor in the fabrication of complex graphite architecture is the design of viscoelastic ink that exhibits essential rheological characteristics conducive to DIW. **Figure 1a** demonstrates the schematic of 3D printing of graphite through DIW method, the nozzle magnified to illustrate the particles' alignment through the tapered tip due to the shear force acting on the graphite flakes. To achieve high-resolution 3D printing, the ink should uniformly extrude through the nozzle without discontinuity and particle jamming during the printing process. We observed two major issues in printing the unmodified graphite slurry - particle jamming in nozzle and separation of graphite and water under pressure. In order to overcome these issues, nano-clay was used as a rheology modifier (ink preparation method in the Supplementary Information) rendering required viscoelastic properties to the ink. Figure 1b-c illustrates the effect on the rheological properties of graphite slurry due to the addition of nano-clay. The unmodified and modified graphite slurry showed a viscosity of 0.15 Pa-s and 2.18 Pa-s at the shear rate of 100 $s^{-1}$, respectively. At a low shear rate (~1$s^{-1}$), the unmodified graphite ink exhibits a viscosity of ~9 Pa-s while the modified graphite ink shows ~345 Pa-s, which is 38 times higher than the unmodified one. Additionally, to facilitate the printing process, graphite ink must exhibit significantly large storage modulus to retain the filamentary shape after extrusion from the nozzle. Figure 1c shows the oscillatory measurement at different strains for modified and unmodified graphite inks. The modified graphite ink exhibits significantly higher storage modulus (G′) compared to the unmodified graphite ink. The incorporation of nano-clay increases the storage modulus of graphite ink by more than 7 times. The modified graphite ink shows the storage modulus of around 90 kPa at very low strain (0.1%), while the unmodified ink displays a storage modulus of 12 kPa. Another key figure of merit, the comparison between viscous and elastic material action, is the relative dissipation – the ratio of G″/ G′, related by phase angle, also known as loss tangent (tan δ). For the modified graphite ink, the loss tangent value is less than unity at low oscillation strain, indicating more solid like (elastic) response of the ink and thus facilitates the filamentary shape retention on exiting the printing nozzle. Figure 1b-c confirmed that the modified graphite ink exhibits rheological properties suitable for DIW. Figure 1d exhibits the high-resolution 3D printed complex graphite structure. This structure was printed using a 1.6 mm tapered nozzle, with the 2.5 cm length, 2.5 cm width, and 2 cm height (approximately 40 layers). The SEM micrograph (Figure 1e) illustrates the resolution of printing in the z-direction. As a result of shear force in the tapered nozzle, the graphite sheets align after extrusion from the nozzle, Figure 1f shows aligned graphite sheets in higher magnification. Further examination of graphite particle motion inside the ink is determined by simulation and analysis of ink flow in the tapered tip. The motion of the liquid is assumed to be a steady laminar flow, and graphite sheets are represented as rigid oblate particles with an aspect ratio $l<1$ (detailed information about liquid motion simulation is available in the method section, Supplementary Information). The motion slows down near the turning points, producing the alignment effect. In the case of oblate particles ($l<1$), the orbits become sharper with decreasing aspect ratio, and the particle spends more time with its smallest dimension in alignment with the direction of the flow. This alignment effect is illustrated in Figure 1g, showing the solution of Jeffery's equations for graphite sheets with an aspect ratio, $l = 0.2$ in a Poiseuille flow with an average velocity of 12 mm/s in a cylinder with 0.8 mm radius. The Figure shows the evolution of graphite sheets starting with random initial orientations and moving along the flow (from left to right) at different radial distances from the center of the cylinder. As shown, the alignment effect is highest near the wall, where shear is maximum, with particles orienting parallel to the flow direction (details about the simulation of liquid motion and particle orientation are available in the method section, and Figure S1 in the Supplementary Information). After printing, the structures were kept at room temperature until the water evaporates to obtain a robust structure. The nano-clay acts as a binder to prevent the disintegration of graphite structures as well as a rheological modifier for direct writing. Figure 1i-k illustrates the energy-dispersive X-ray spectroscopy (EDX) results, demonstrating the role of nano-clay. The presence of carbon and silicon elements (Figure 1i and j) corresponds to graphite and clay in the printed construct. The overlay of these elements (Figure 1k) indicates the uniform distribution of nano-clay in the graphite ink, which gets attached to the graphite layers and binds them together, yielding a self-supporting solid structure.

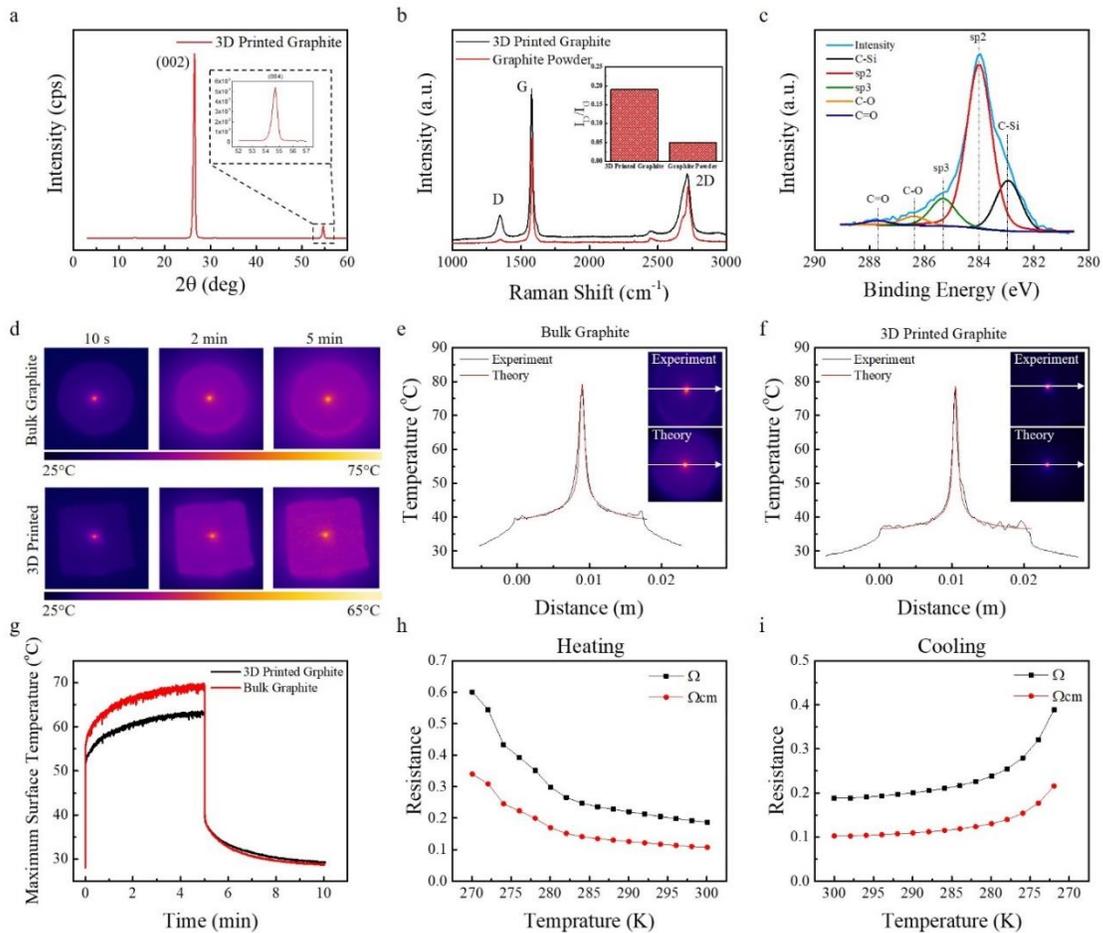

**Figure 2. Structural, thermal, and electrical properties of 3D printed graphite.** a) XRD spectra of 3D printed samples; b) Raman spectra of graphite powder and printed samples. c) XPS spectra of 3D printed graphite. d) Temperature mapping of the bulk graphite and 3D printed graphite surface using IR camera during illumination (450 nm wavelength) in air. e and f) Comparison between experimental and computational temperature profiles in the same region

(indicated with a white arrow in the temperature mapping shown in inset) of bulk graphite and 3D printed graphite, respectively. g) Temporal profile of the maximum surface temperature of bulk and 3d printed graphite during illumination of laser in the air. h and i) The electrical resistance of 3D printed samples at different temperatures during the heating and cooling process.

Structural characterization was carried out on the printed graphite structure to illustrate the effect of the printing process. **Figure 2**a shows the X-ray diffraction (XRD) pattern of 3D printed graphite structure. We observed two characteristic peaks at $2\theta = 26.7°$ and $2\theta = 54.7°$, which represents the Bragg reflections of the (002) and (004) planes, respectively. The absence of diffraction lines other than the (00l) lines demonstrates the high degree of alignment of the graphite flakes[16,17], as shown in the SEM images in Figures 1e and 1f. Moreover, the fact that the intensity of the (004) reflection is much lower than that of the (002) reflection implies that there are large twist angles between the graphite planes.[18] Such twisting is likely to occur during the restacking of the graphite flakes during extrusion through the nozzle (Figure 1g), promoting the formation of turbostratic carbon (or turbostratic graphite). Turbostratic carbon is a form of graphite in which the stacking of the graphene layers differs from the thermodynamically favorable AB stacking found in pure graphite[19,20]. In addition to the restacking of graphite flakes during the printing process, the formation of turbostratic graphite could stem from the ball milling process used for the preparation of the graphite ink. It has been reported in the literature that the ball milling can induce enough energy to stimulate the formation of the turbostratic structure.[19,21]

Raman spectroscopy measurement (Figure 2b) shows three primary distinguishable vibrations for the 3D printed graphite and the pristine graphite powder: (i) The vibration that is represented by the band at ~1350 cm$^{-1}$, known as D-band (defects), is related to a disordered graphitic structure with an $A_{1g}$-symmetry (ii) The Raman active vibration at ~ 1580 cm$^{-1}$ corresponds to a highly ordered graphite-like structure related to an $E_{2g}$-symmetry, known as G-band (in-plane C-C stretching vibration) and (iii) vibration at around ~2850 cm$^{-1}$, known as 2D band, which is a second-order contribution of the D band, involving a two-phonon process.[22] As can be observed from Figure 2b, the two main differences in the Raman spectra are- (1) the change in the line shape of the 2D band and (2) the $I_D/I_G$ ratio (intensity of D band to intensity of G band ratio). The explanation for the first consists mainly in the formation of turbostratic carbon, as discussed previously. The lack of AB stacking in turbostratic graphite results in electronic decoupling and a weaker interaction between the layers which, in turn, is evidenced by a single Lorentzian fit to the 2D band as a spectroscopic signature.[20,23,24] Another contribution for the single Lorentzian fit to the 2D band can arise due to the broadening of the 2D band that occurs because of the reduction of the graphite crystallite size during the ball milling process.[25] The increase in the $I_D/I_G$ ratio also traces back to the initial ball milling process and the reduction of the average crystallite size followed by the introduction of structural defects in the graphite flakes during the milling process. This results in the increase of the D band whereas the G band is not much affected by the introduction of defects.[22–24]

In order to determine the elemental composition of the 3D printed sample, X-ray photoelectron spectroscopy (XPS) was performed. Figure 2c presents a high-resolution XPS spectrum of the printed graphite (the survey spectrum shown in Figure S4 in the Supporting Information). The spectrum of C 1s is deconvoluted in a total of 5 peaks: The peak at 284 eV is assigned to $sp^2$ C=C bond, the peak at 285.32 eV is assigned to $sp^3$ C−C, and the remaining peaks are attributed to C−Si bonds at 282.94 eV, C−O groups at 286.38 eV, and C=O groups at 287.69 eV. Except for the C−Si peak, the obtained deconvoluted peaks are very similar to those obtained for ball-milled graphite powder.[26] In this case, the C−Si peak (the peak at ~283 eV) accounts for the bonds between the graphite flakes and the nano-clay added in the graphite ink, since nano-clays are mostly silicates.[27] The peaks at 286.38 eV and 287.69 eV can be attributed to C-O and

C=O as residual contamination from the air.[28] Although a pure graphitic system should mostly contain $sp^2$ bonds, it was demonstrated that the $sp^3$ peak in graphitic systems arises due to defects.[26,29] In the printed constructs, possible sources of defects that can potentially disturb the $sp^2$ configuration are the C−Si bonds, oxygen contamination from the air and defects generated in the graphite flakes due to the milling process, which lead to C−H bond termination, promoting a transition of the $sp^2$ carbon to the $sp^3$ configuration.[26]

Graphite is widely used in refractory foundries and heat sink applications; it is of practical importance to investigate the effect of the printing process on the thermal properties of the 3D printed structures and compare it with bulk samples. Figure 2d presents the high-resolution spatial and temporal mapping of temperature variations of bulk graphite, and 3D printed graphite under illumination. The thermal conductivity of both samples was obtained through indirect measurement and leveraging from the thermal simulation of the experiment (Figure 2e and 2f, and details about thermal simulation are available in the method section, Figure S2-S3, and Table S1 in Supplementary Information). To determine the thermal conductivity, the single line temperature profile was extracted from temperature mapping captured by the IR camera for both samples. Then the thermal conductivity of printed samples and bulk graphite was found out by fitting the experimental temperature profile to the simulation in the same region (indicated by a white arrow in the inset images of Figures 2e and 2f). As shown in Figures 2e and f, the temperature profiles are in good agreement in experiment and simulation yielding almost equal thermal conductivity for the 3D printed sample (~9 W/m-K) and bulk graphite (10 W/m-K)[30]. Thus, the printing process and ink additives do not affect the heat transfer behavior of the graphite structure. Figure 2g shows the temporal profile of the maximum temperature of the surface during the illumination of the laser and the cool-down rate after removing the illumination source. Interestingly, the heat dissipation rate has been found to be very similar for both samples (Figure 2g). The differences in the maximum surface temperature of 3D printed and bulk graphite during the illumination period, though negligible, suggest that the 3D printed sample has slightly more heat resistance, which is likely due to the addition of nano-clay.[31] Clay addition is also expected to impart better thermal stress distribution in the printed samples. As can be seen from Figure 2d, for the bulk graphite samples, a higher temperature is observed during illumination along the circumferential edges, indicating local accumulation of thermal stresses. On the other hand, the temperature profile is more evenly distributed for the corresponding 3D printed samples, suggesting improved thermal distribution. The electrical conductivity of the graphite structure is one of the critical elements in many applications like electrical circuits, heater element, and electrodes. In this study, the electrical resistance was determined at different temperatures during the heating and cooling of the 3D printed graphite sample (Figures 2h and i, respectively). The average resistivity of the 3D graphite structure is ≈ 0.18 $\Omega$ cm. Although clays are essentially insulating materials and hence, would act as potential barriers in the electronic conduction path in the printed sample, the small dimension of clay platelets and therefore, smaller barrier width, might allow the electrons to quantum mechanically tunnel through them and maintain almost similar conductivity as that of pure graphite.[32]

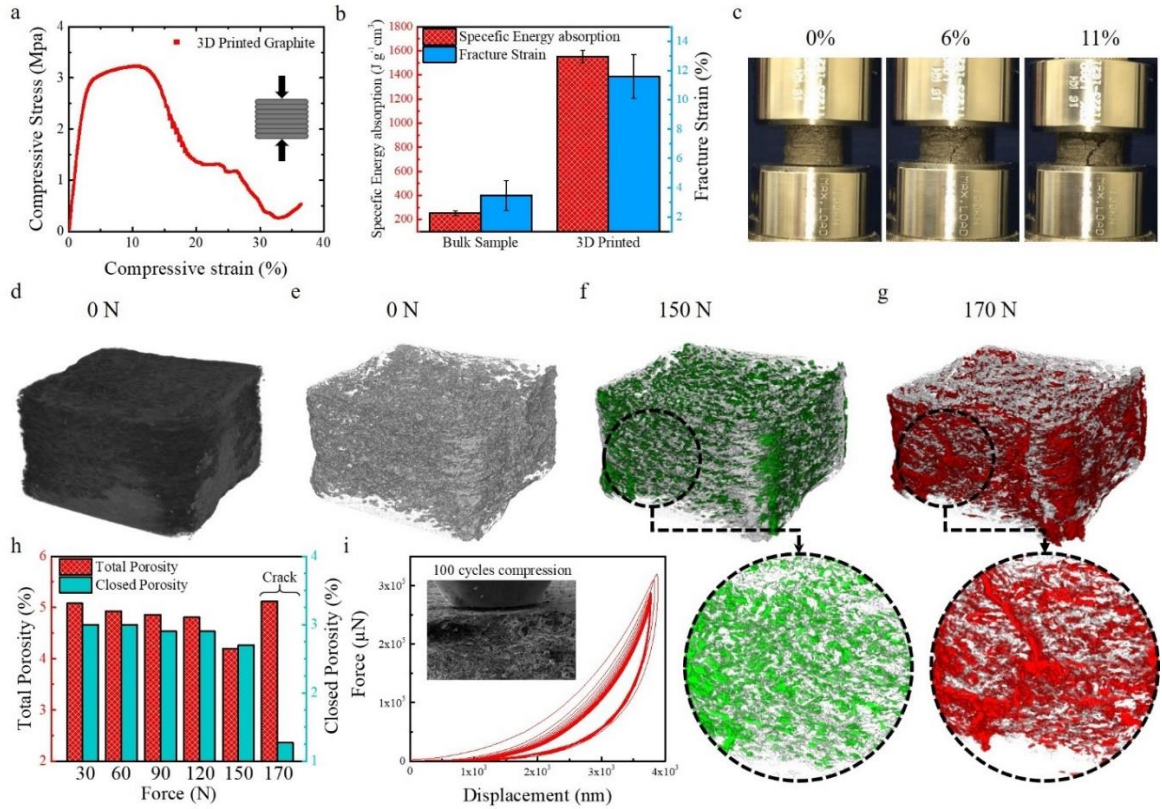

**Figure 3. Mechanical properties of 3D printed graphite.** a) The compressive stress-strain curve of 3D printed graphite structures from the uniaxial compression test by applying load perpendicular to the printing path. b) The specific energy absorption and fracture strain for 3D printed graphite and bulk graphite sample. c) Deformation mechanism of 3D printed graphite structure under compressive load at different stain levels. d) CT scan image of the printed structure before applying load. e-g) Demonstration of pores distribution and movement in the printed graphite under various load levels (grey, green, and red contour indicate 0, 150, and 170N, respectively). Magnified images illustrate the pores' movement and crack initiation by merging closed pores upon increasing applied load. These micro-CT images were measured at 15µm pixel resolution. h) The total and closed porosity percentage under various load levels. i) Force-displacement measurement of 3D printed graphite under microscale dynamic compressive loading.

Graphite is considered as a brittle material in the macro-scale, which can undergo catastrophic failure at low strain (less than 2%) and shows poor fracture-toughness properties. Previous studies have revealed that topology plays a crucial role in the mechanical properties of materials, and the realization of complex topologies such as Schwarzites, Tubulanes, and honeycomb patterns leads to the improvement of structural toughness.[33–35] Thus, the enhancement of mechanical properties of graphite structure is possible by printing complex topologies using additive manufacturing techniques. Since DIW printing is a layer by layer process, the printed structure has an anisotropic property, and the mechanical properties depend on the printing path and direction. Therefore, the uniaxial compression test was carried out on the 3D printed graphite structure perpendicular and parallel to the printing direction, and the related stress-strain curves are shown in Figure 3a and Figure S5. As can be seen, the printed graphite structure exhibits ductile behavior when the load is applied perpendicular to the printing path. Atypical for a brittle material like graphite, the printed graphite structure first experiences a linear (elastic) region for compression strain up to 3%, then a plateau (plastic) region beyond 14% strain followed by a nonlinear (failure) region. However, this was expected from the viscoelastic characterization of the modified ink (Figure 1c), which exhibited enhanced loss modulus throughout the entire window of the experimental timescale (strain rate in this case), suggesting enhanced energy dissipation mechanism in the modified ink than the unmodified one. The graphite sheets after extrusion from the nozzle facilitate the dislocation and movement of graphite sheets in the plane and enhance the fracture strain for the compression test carried out perpendicular to the printing path. The alignment of the printed graphite sheets is further confirmed by the compression test carried out in the direction of the printing path, which exhibits a progressive crushing phenomenon- a characteristic of buckling failure (Figure S5, Supplementary Information). In Figure 3b, the specific energy absorption and fracture strain were compared for printed and bulk graphite, which was shown to increase by more than 6 and 3 times, respectively. The deformation mechanism of 3D printed graphite cylinder is shown in a series of snapshots at different strain rates, which confirms the ductile behavior of the 3D printed graphite structure. Interestingly, the printed samples do not disintegrate even at 11% strain, while the bulk samples collapse at 2% compressive strain (Figure S6, Supplementary Information). The crack occurred along the shear propagation direction, which could be attributed to the elimination of layer boundary and form a strong interaction between layers in our printing method. To better understand the failure mechanism of 3D printed graphite structure, the *in situ* micro-Computed Tomography (CT) scan images were captured at different compressive loads. The porosity of the 3D printed structure was obtained through high resolution (800 nm pixel) CT scanning (Figure S7, Supplementary Information). Figure 3d shows the CT images of 3D printed nonpatterned graphite in the initial stage, and Figure 3e represents the pores in the same structure. To demonstrate the crack initiation and propagation phenomena, the distribution and movement of the pores under different compressive loads are shown in Figure 3e-g. The grey color represents the initial condition of the structures, and the color shift from green to red indicates the increase in the compressive load. While applying the compressive load on the graphite sample, we observed the movement of pores; by increasing the load level, the small pores connected and initiated crack. The fastening of the pores and consequent crack initiations were evident in the magnified CT images of Figures 3f and g. The total porosity (summation of closed and open pores) and closed porosity (the pores that do not have a connection with the outside of the sample) were presented as a function of the applied force in Figure 3h. By increasing the force, the closed pores percentage decreased since they densified and connected to form open pores, which have a connection with the outside of the sample, thus yielding cracks. Before crack initiation, the percentage of total porosity decreased, and right after crack formation, it increased rapidly.

Figure 3i presents the 100 compressive cycles of force-displacement curves on the printed graphite sample, which exhibits a somewhat consistent hysteresis during multiple loading and unloading events. Generally, hysteresis in force-displacement curves suggests that there is energy dissipation, and the decrease in the force-displacement hysteresis after the first or first few loading-unloading cycles is indicative of lower fracture energy. However, in our case, the repeatable hysteresis over multiple loading-unloading cycles suggests that the printed structure has high fracture energy, and is anti-fatigue at microscale deformations.[36]

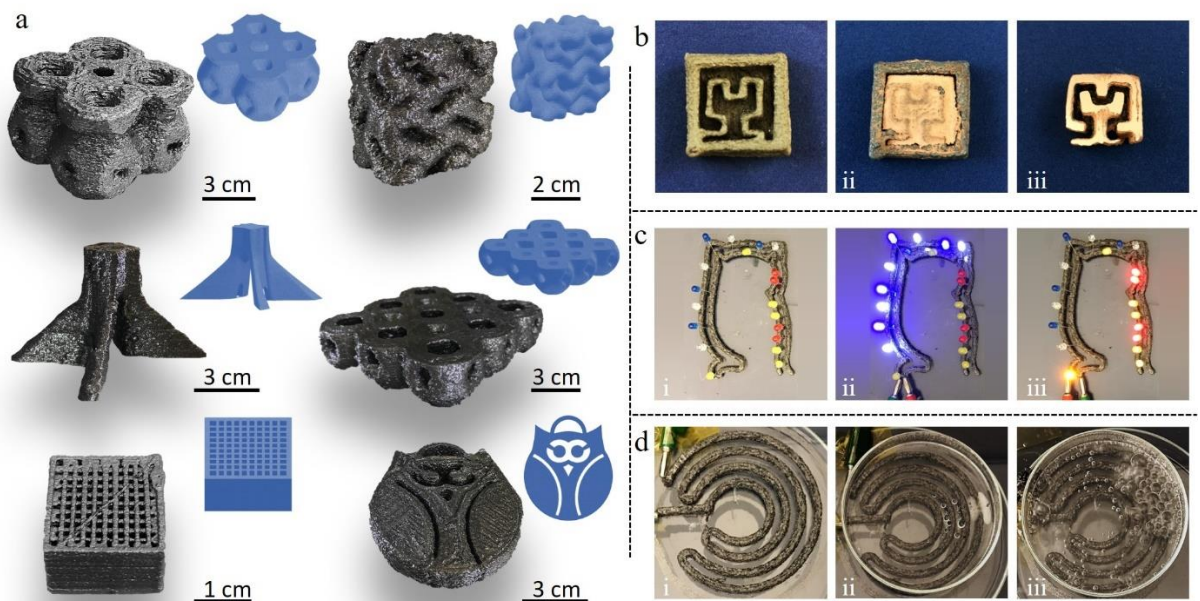

**Figure 4. Applications of 3D printed graphite.** a) 3D printed complex structures with modified graphite ink. b) The 3D printed casting mold of graphite ii) Graphite mold filled with copper powder. iii) the copper piece after sintering in the furnace. c) i) The printed graphite circuit in the shape of an owl, which has several LED bulbs. ii) the blue LEDs were turned on by applying the DC voltage, iii) by changing the voltage polarity red LEDs (leads are connected vice versa for blue and red LEDs) were turned on. d) The printed heater element, ii) the petri dish full of water placed on top of the heater. iii) By applying an appropriate power, the water temperature reached the boiling point.

The high plasticity, electrical and thermal conductivity of the 3D printed graphite structures have been utilized in a series of applications, as shown in Figure 4. Graphite is used as a mold for metal casting; however, due to the brittle nature of graphite, fabricating a complex-shaped mold has been a traditionally challenging task. The additive manufacturing technology can significantly expand the production of complex topology mold, which remains unattainable with the conventional synthesis method. Figure 4a shows an array of structures, which were 3D printed using the graphite ink, demonstrating the feasibility of achieving complex shapes. The top two structures in Figure 4a represent theoretical structures called Schwarzite, which has positive and negative Gaussian curvatures and could not be built without additive manufacturing technology.[34] These structures were printed using 1.64 mm tapered nozzles of different sizes. The height of these structures ranges from 1cm to 5 cm (20 to 100 layers, respectively). This confirms the high stiffness of the graphite ink, enabling the printed construct to withstand the load of the above layer(s) without deformation. Figure 4b demonstrates the 3D printed graphite mold (I), the mold filled with copper powder (II) and the copper part after sintering in the furnace (III) which has the exact shape of the mold (more information related to casting metal with different methods and elements are available Figure S8 in the Supplementary Information which reveals the durability of graphite mold). Figure 4b shows a printed graphite circuit in the shape of an owl, which has several LED bulbs connected to it. Figure 4c II and III show the LEDs turn on by applying a 3V DC voltage on the circuit, and by changing the positive and negative pole, different colors of LEDs were turned on. Another critical application of graphite is manufacturing 3D elements for heating applications. Figure 4d (I) shows the printed graphite element connected to a power supply. A petri dish full of water was placed on top of the graphite heater element (Figure 4c II). Once the appropriate voltage was applied, the water started boiling (Figure 4d III). In all of these applications, graphite can be printed in different complex forms, which possess excellent electrical and thermal conductivity and can be used right after printing without any further processing like sintering.

In summary, we demonstrate the development of a graphite ink with essential rheological properties conducive to direct ink writing of complex 3D graphite architecture. Electron microscopy studies revealed shear-induced alignment of graphite sheets, further analyzed, and confirmed by simulation of the ink flow and mechanical testing as well. Experimental and computational temporal temperature profiling revealed excellent heat transfer behavior of the 3D printed sample. The printed graphite structures have superior electrical properties, as shown by the IV measurements and mechanical properties confirmed by uniaxial compression test, force-displacement measurement, and CT scan imaging. Finally, the structures were successfully incorporated in a series of applications, demonstrating the power of the DIW technique in the design of practical and intricate large-scale multifunctional 3D graphitic architectures.

## Acknowledgment

The authors acknowledged Aramco Research Center for funding this research (Grant number: 1137681). L.M.S. acknowledges CAPES (Coordination for the Improvement of Higher Education Personnel) under the Brazilian Ministry of Education for the financial support.

## Author contribution

S.M.S, M.M.R., C.S.T., P.M.A conceived and coordinated the research, S.M.S and C.D processed the sample and performed the mechanical testing. S.B performed Nanomechanical test. L.V, A.K and R.V conducted CT scanning and reconstructed the data for quantitative analysis. S.M.S, S.E and L.M.S conducted material characterization. A.A, M.L and M.M.R. did the thermal conductivity and related analysis, S.M.S and A.B.P performed electrical conductivity testing. S.M.S, C.S.T and R.V investigated the structure, all author analysed and discussed the results, S.M.S, L.M.S, P.B, C.T, C.S.T, M.M.R and P.M.A wrote the manuscript.

## Additional information

Method

Figures S1-S8

Table S1

## Competing financial interests

The authors declare no competing financial interests.